\documentclass[aps,prL,twocolumn,superscriptaddress]{revtex4-1}
\usepackage{graphicx}
\pdfoutput=1

\usepackage{dcolumn}
\usepackage{bm}
\usepackage{amsmath}
\usepackage[english]{babel}
\usepackage[utf8]{inputenc}
\usepackage{verbatim}
\usepackage{natbib}

\usepackage{layouts}

\begin{document}
	\title{A two-dimensional array of single-hole quantum dots}
	
	\author{F. van Riggelen}
	\email{f.vanriggelen@tudelft.nl}
	\affiliation{QuTech and Kavli Institute of Nanoscience, Delft University of Technology, Lorentzweg 1, 2628 CJ Delft, The Netherlands}
    \author{N. W. Hendrickx}
	\affiliation{QuTech and Kavli Institute of Nanoscience, Delft University of Technology, Lorentzweg 1, 2628 CJ Delft, The Netherlands}
	\author{W. I. L. Lawrie}
    \affiliation{QuTech and Kavli Institute of Nanoscience, Delft University of Technology, Lorentzweg 1, 2628 CJ Delft, The Netherlands}
	\author{M.~Russ}
	\affiliation{QuTech and Kavli Institute of Nanoscience, Delft University of Technology, Lorentzweg 1, 2628 CJ Delft, The Netherlands}
	\author{A. Sammak}
	\affiliation{QuTech and Netherlands Organization for Applied Scientific Research (TNO), Stieltjesweg 1 2628 CK Delft, The Netherlands}
	\author{G. Scappucci}
	\affiliation{QuTech and Kavli Institute of Nanoscience, Delft University of Technology, Lorentzweg 1, 2628 CJ Delft, The Netherlands}
	\author{M. Veldhorst}
	\email{m.veldhorst@tudelft.nl}
	\affiliation{QuTech and Kavli Institute of Nanoscience, Delft University of Technology, Lorentzweg 1, 2628 CJ Delft, The Netherlands}
		
	\begin{abstract}
    Quantum dots fabricated using techniques and materials that are compatible with semiconductor manufacturing are promising for quantum information processing. While great progress has been made toward high-fidelity control of quantum dots positioned in a linear arrangement, scalability along two dimensions is a key step toward practical quantum information processing. Here we demonstrate a two-dimensional quantum dot array where each quantum dot is tuned to single-charge occupancy, verified by simultaneous measuring with two integrated radio frequency charge sensors. We achieve this by using planar germanium quantum dots with low disorder and small effective mass, allowing the incorporation of dedicated barrier gates to control the coupling of the quantum dots. We demonstrate hole charge filling consistent with a Fock-Darwin spectrum and show that we can tune single-hole quantum dots from isolated quantum dots to strongly exchange coupled quantum dots. These results motivate the use of planar germanium quantum dots as building blocks for quantum simulation and computation.
	\end{abstract}
	\maketitle
	 
	Quantum information requires qubits that can be coherently controlled and coupled in a scalable manner \cite{DiVincenzo2000}, while quantum error correction and scalable interconnects demand the ability to couple qubits along at least two dimensions \cite{Terhal2015,Franke2019}. Across all the different qubit technologies, quantum dots \cite{Loss1998} fabricated using techniques compatible with standard semiconductor manufacturing are particularly promising \cite{Vandersypen2017}. Furthermore, realizing two-dimensional quantum dot arrays may allow to construct highly scalable qubit tiles such as crossbar arrays \cite{Li2018} supporting quantum error correction \cite{Helsen2018} for fault-tolerant quantum computation.
	
	A key challenge is therefore to develop two-dimensional arrays of quantum dots that exhibit a high level of uniformity, long quantum coherence, and that can be operated with excellent control. Initial research centered around low-disorder gallium arsenide (GaAs) heterostructures \cite{Petta2005, Koppens2006}, which advanced to exciting demonstrations such as coherent spin transfer across an array of quantum dots \cite{Kandel2019}, and the operation of a two-dimensional quantum dot array \cite{Mukhopadhyay2018}. Nonetheless, group III-V materials suffer from hyperfine interaction, resulting in fast spin dephasing and reduced operation fidelity. Instead, group IV materials can be isotopically enriched \cite{Itoh1993, Itoh2014} to virtually eliminate dephasing due to a nuclear spin bath. This has stimulated research on silicon and led to orders of magnitude improvement in coherence times \cite{Veldhorst2014, Muhonen2014}. While advances in devices based on silicon heterostructures have led to the operation of linear arrays containing up to nine quantum dots \cite{Zajac2016}, the relatively large effective mass of silicon electrons, the presence of valley energy states, and the finite disorder complicates progress \cite{Zwanenburg2013}. Though fabrication is advancing to complementary metal–oxide–semiconductor (CMOS) foundry-manufactured devices \cite{Maurand2016, Pillarisetty2018}, demonstrations on two-dimensional quantum dot arrays have been limited to reaching single-electron occupancy in a triple quantum dot \cite{Ansaloni2020, Chanrion2020, Gilbert2020}. Reaching simultaneously the single-charge regime with all quantum dots in a two-dimensional array fabricated using CMOS foundry compatible materials remains thereby an outstanding challenge.

	\begin{figure*}
	\includegraphics[width = \textwidth]{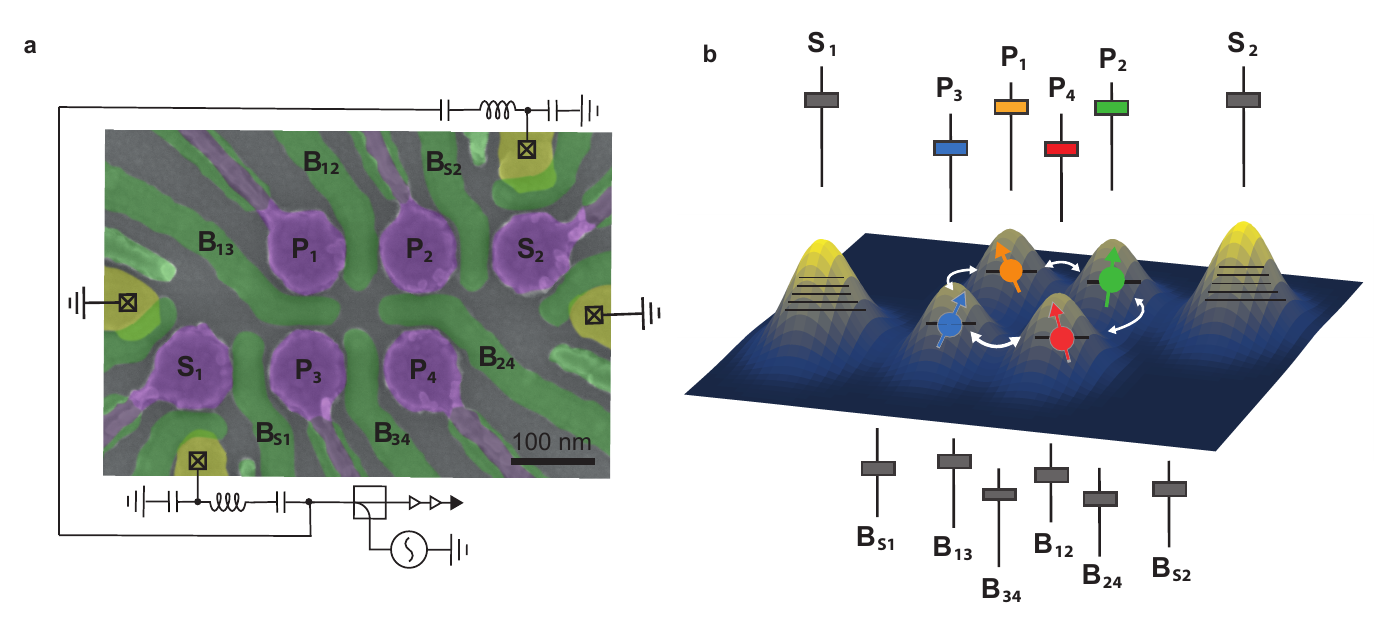}%
	\caption{\textbf{A 2x2 germanium quantum dot grid with two integrated rf sensors} (a) False colored SEM image of a sample similar to the one on which the measurements are performed. The plunger gates of the quantum dots P are colored in purple, the barrier gates B are colored in green and the aluminum ohmics in yellow, which serve both as source and drain contacts for rf sensing, as well as charge reservoirs for the quantum dots. (b) Schematic representation of the potential landscape, illustrating how the plunger and barrier gates control the quantum dots. In the image, each quantum dot is occupied with a single hole (N=1), which is color coded per quantum dot (yellow for $Q_1$, green for $Q_2$, blue for $Q_3$, and red for $Q_4$). The charge occupation in a quantum dot is controlled by a plunger gate, symbolized by a slider above the image with the same color. The sensing dots are tuned into the multi-hole regime, illustrated by the many energy levels drawn in the quantum dot. The coupling between the quantum dots, indicated by the arrows, is controlled by a barrier gate, depicted by a slider below the image.}
	\label{fig:4quantum dotsample}
    \end{figure*}

	Germanium is rapidly emerging as an alternative material to realize spin qubits \cite{Scappucci2020}, since holes in germanium have favorable properties such as a small effective mass \cite{Lodari2019}, large excited states due to the absence of valley degenerate states, and strong spin-orbit coupling for electrically driven single-qubit rotations without the need for external components \cite{Bulaev2005, Bulaev2007, Watzinger2018}. High-quality Ge/Si core-shell nanowires enabled the construction of a triple quantum dot in a linear arrangement, albeit only in the multi-hole regime \cite{Froning2018}. The realization of high-quality strained Ge/SiGe quantum wells \cite{Sammak2019} has led to the development of quantum dots \cite{Hendrickx2018, Lawrie2020}, demonstration of long hole spin relaxation times \cite{Lawrie2020b}, the operation of a single-hole qubit \cite{Hendrickx2019b}, and enabled the execution of two-qubit logic in germanium \cite{Hendrickx2020}. Furthermore, quantum dots in planar germanium are realized using industry compatible techniques \cite{Pillarisetty2011}, promising large-scale implementations provided that germanium quantum dots can be engineered beyond linear arrangements.

	Here, we realize a two-dimensional quantum dot array using materials compatible with existing CMOS technology and demonstrate a quadruple germanium quantum dot. We obtain excellent control over the charge occupancy and the interdot coupling. The device consists of the quantum dot grid and an additional two quantum dots on the sides that are used for radio frequency (rf) charge sensing. We are able to tune each quantum dot to the single-hole occupancy and we find shell filling to be consistent with a Fock-Darwin spectrum. This demonstrates a qubit state manifold with large separation energy, since excited states, such as valley energy states, are absent. We exploit the integrated barrier gates to gain independent control over the hole occupancy and the tunnel coupling between neighboring quantum dots. We use this to demonstrate the single-hole occupancy in the full quadruple quantum dot array as a stepping stone toward two-dimensional arrays of quantum dot qubits.

	\begin{figure*}
	\includegraphics[width = \textwidth]{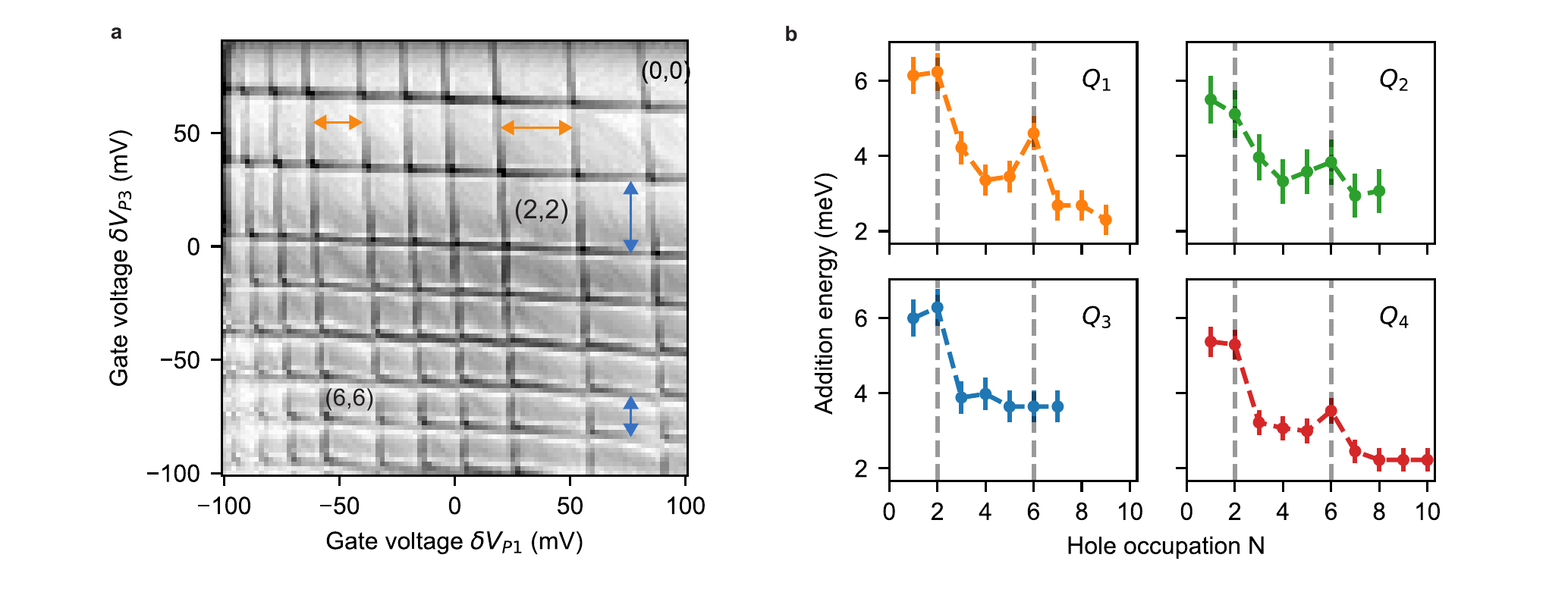}%
	\caption{\textbf{Charge filling in the individual quantum dots} (a) Shown is a charge stability diagram of the double quantum dot $Q_1$ - $Q_3$ with negligible tunnel coupling (See Supplementary Information section I for the quantum dot pair $Q_2$-$Q_4$). Here, the results are shown as measured with sensor $S_1$, the sensor closest to the quantum dot pair. We can observe all transitions with both sensors, albeit with reduced sensitivity for the more remote quantum dots, as shown in Figure \ref{fig:4quantum dotCSD_first_try}. The hole occupation ($N_{Q1}$, $N_{Q3}$) is indicated in the charge stability diagram. (b) Addition energy for the four quantum dots, extracted from the corresponding stability diagrams and converted using a lever arm $\alpha =$ 0.19 eV/V. The dashed grey lines correspond to the hole fillings for which increased addition energy is expected due to shell filling when considering a circular potential landscape and spin degeneracy (also indicated by orange and blue arrows in (a) for $Q_1$ and $Q_3$ respectively).}
	\label{fig:Shell_filling3}
    \end{figure*}
	
	Figure \ref{fig:4quantum dotsample}a shows a scanning electron microscopy (SEM) image of a quantum dot grid and Figure \ref{fig:4quantum dotsample}b shows a schematic image of the potential landscape and the control gates of the quantum dot system. Fabrication is based on a multilayer gate design \cite{Lawrie2020}. Holes in strained germanium benefit from a low effective mass, low disorder, and absence of valley states. These assets ease constraints in fabrication and relax the quantum dot design, which makes it possible to define a 2x2 quantum dot grid with only two overlapping gate electrodes. The quantum dots are defined using plunger gates P and are coupled through barrier gates B. We have fabricated the barrier gates as the first layer and the plunger gates as the second layer, which results in a good addressability of both the tunnel couplings and quantum dot energy levels. The aluminum ohmics serve as hole reservoirs for the charge sensors. Controllable loading of the quantum dots is obtained through an additional barrier gate between the sensor and the quantum dots ($B_{S1}$ and $B_{S2}$). The charge occupation in the dots is measured with the nearby sensing dots. We use rf reflectometry to achieve a high measurement bandwidth of the sensor impedance, which allows for measuring charge stability diagrams in real time. 
	
		 \begin{figure*}
	\includegraphics[width = \textwidth]{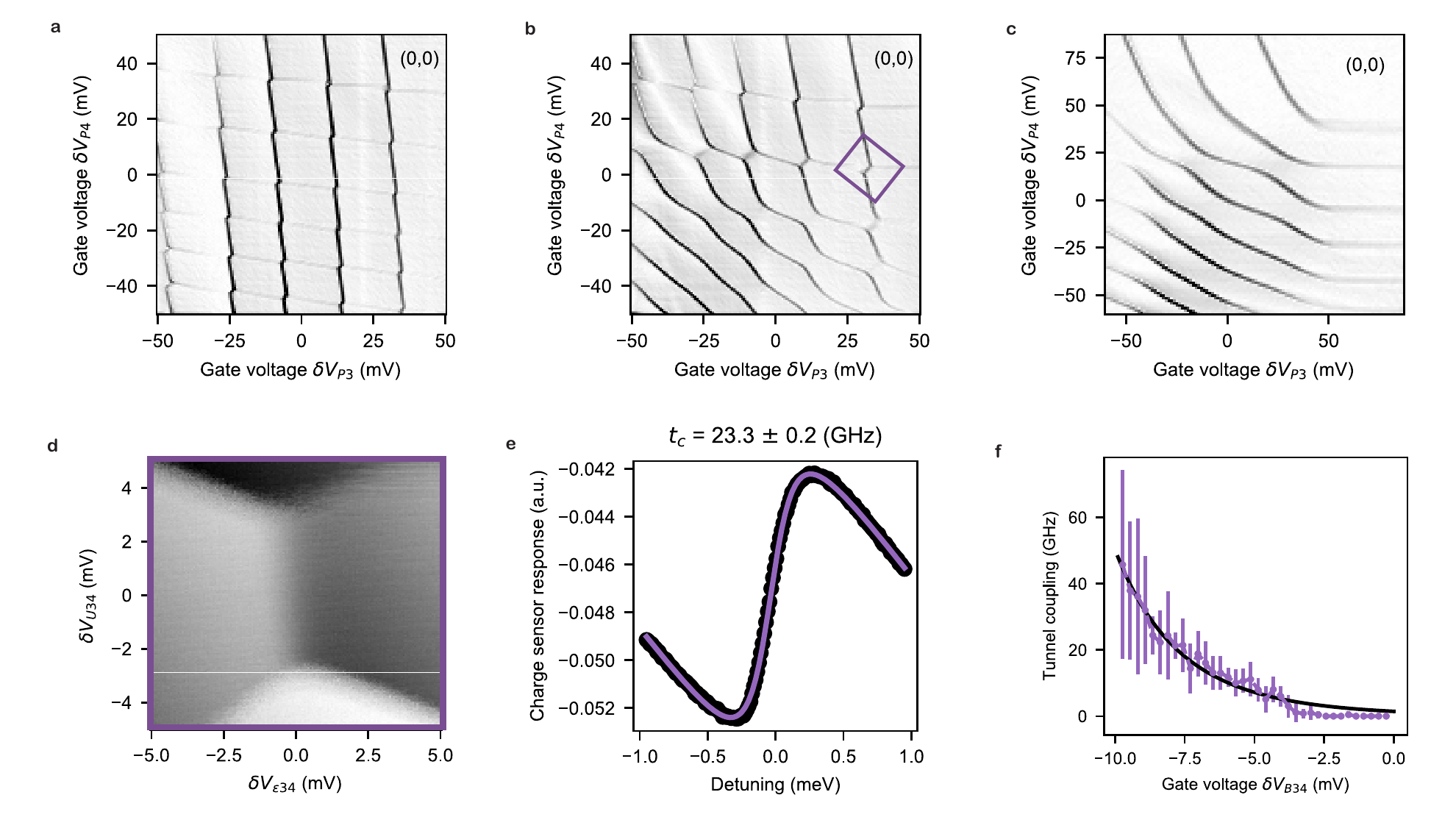}%
	\caption{\textbf{Controllable interdot tunnel coupling.} (a,b,c) Charge stability diagram for quantum dot pair $Q_3$ - $Q_4$  with barrier gate voltage $V_{B34}$ = -1010.6 mV (a), $V_{B34}$ =  -1055.1 mV (b), and $V_{B34}$ = -1137.1 mV (c). By varying the barrier gate voltage we can freely tune the tunnel coupling over a large range. (d) Zoom-in on the relevant (1,1)-(0,2) charge configuration where we quantify the tunnel coupling. (e) By fitting the charge polarization line \cite{DiCarlo2004} we obtain the tunnel coupling, which is $t_C$ = 23.3 $\pm$ 0.2 GHz. (f) By varying the gate voltage $V_{B34}$ we can control the tunnel coupling up to 40 GHz. Reduced charge sensor sensibility for higher tunnel coupling causes the uncertainty in the measurement to increase. The trend of the tunnel coupling corresponds well to a fit based on the WKB theory (see Supplementary Information section IV for further details).}
	\label{fig:tunnelcoupling}
    \end{figure*}
    
       \begin{figure*}
	\includegraphics[width = \textwidth]{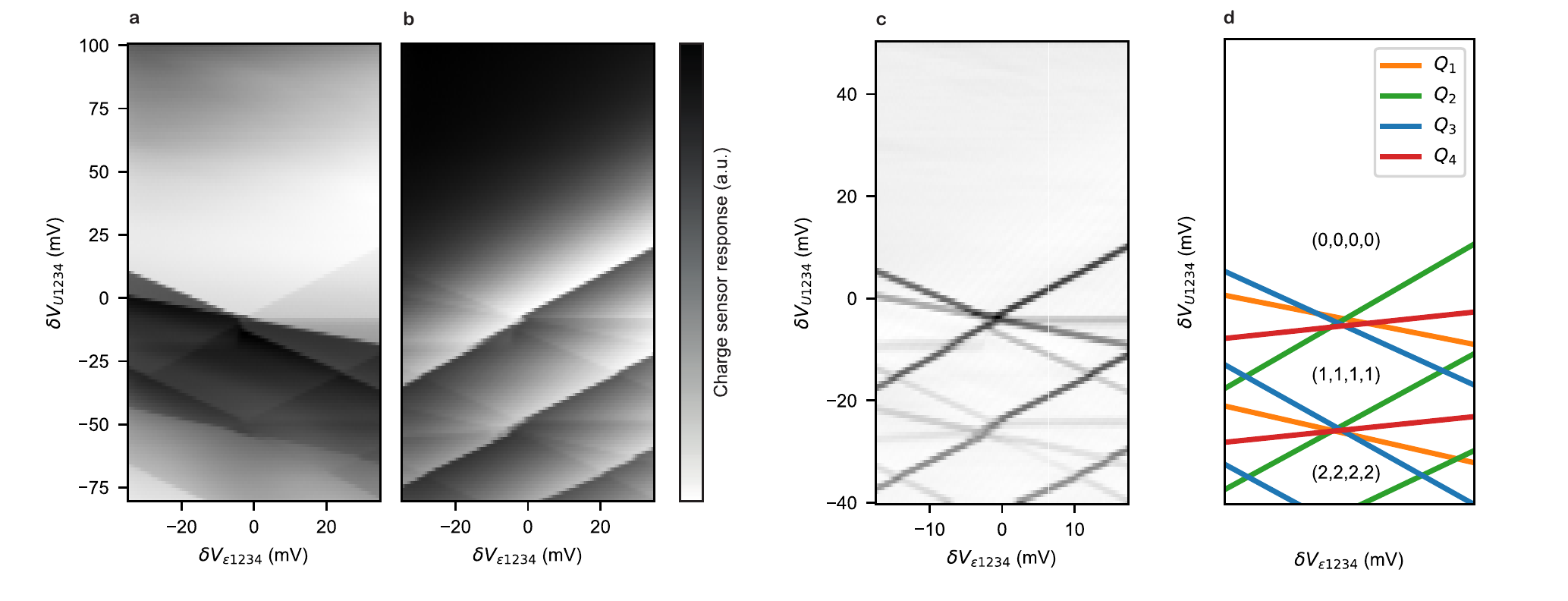}%
	\caption{\textbf{Quadruple quantum dot in germanium} (a,b) Charge stability diagram of the four quantum dot system, obtained by simultaneous readout of $S_1$ and $S_2$. (a) Charge sensor response of sensor $S_1$. (b)  Charge sensor response of $S_2$. While we can observe all transitions with each sensor, we observe a significant larger sensitivity to the quantum dots neighboring the sensor. (c) Derivative of the combined response signal, clearly revealing the charge addition lines for each of the quantum dots. (d) Schematic representation explaining the charge addition lines as measured in (c), confirming the absence of additional lines from spurious quantum dots or traps and demonstrating a single-hole quadruple quantum dot array. Hole occupation in the dots ($N_{Q1}$, $N_{Q2}$, $N_{Q3}$, $N_{Q4}$) is indicated for an empty system, single-hole occupation, and double hole occupation for all four dots.}
	\label{fig:4quantum dotCSD_first_try}
    \end{figure*}

	Figure \ref{fig:Shell_filling3}a shows a charge stability diagram corresponding to quantum dot pair $Q_1$-$Q_3$. See Supplementary Information section I for the stability diagram corresponding to quantum dot pair $Q_2$-$Q_4$. In this measurement we preserve the sensitivity of the sensor, by offsetting the effect of a change in voltage on the plunger gate of the quantum dots with a small change in voltage on the plunger gate of the sensors. From the linear charge addition lines in Figure \ref{fig:Shell_filling3}a we infer that the capacitive coupling between the plunger gate and the neighboring quantum dot is small and does not require compensation. In Figure \ref{fig:Shell_filling3}b, we show the addition energies for each of the four quantum dots in the few-hole regime. The addition energies are extracted from the charge stability diagrams, by analysing the spacing between the addition lines for all the dots. The change in gate voltage is converted into energy, using a lever arm $\alpha = $ 0.19 eV/V. Steps are observed for hole occupations $N$ = 2 and $N$ = 6 that are consistent with shell filling for a circular quantum dot and considering the spin degree of freedom \cite{Tarucha1996}. These experiments also highlight the absence of low-energy excited states such as valley states, which would give rise to a different shell filling pattern \cite{Lim2011}. It is interesting to observe that quantum dot $Q_1$ and $Q_4$ show shell filling as expected of circular quantum dots, while for $Q_2$ and $Q_3$ the expected peaks in addition energy are less pronounced. Moreover, $Q_2$ and $Q_3$ show an increased addition energy for $N$ = 4. We ascribe this difference to $Q_2$ and $Q_3$ being positioned closely to the sensors quantum dots, which are operated using relatively large negative potentials. The electric field from the sensors might distort the circular potential to a more elliptical shape, which would in turn modify the electronic structure and cause an increased addition energy at half-filling \cite{Reimann2002}.
	
	Having shown control over the hole occupation of the individual quantum dots, we focus on the interdot tunnel coupling. Figures \ref{fig:tunnelcoupling}a-c show charge stability diagrams of a double quantum dot defined by plunger gates $P_3$ and $P_4$ for different barrier gate potentials, compensating the effect of the change in voltage on the sensor. We find that we can tune the quantum dots from being fully isolated, to a strongly coupled regime, and to merging quantum dots,  indicating a high level of tunability. Importantly, we reach all regimes while freely choosing the hole occupancy.
    
    To quantify the tunnel coupling between the quantum dots we analyze the charge polarization lines. Figure \ref{fig:tunnelcoupling}d shows the anticrossing corresponding to the (1,1)-(0,2) charge configurations. We measure charge sensor response along the detuning axis and determine the tunnel coupling by fitting the charge polarization lines \cite{DiCarlo2004}, as shown in Figure \ref{fig:tunnelcoupling}e. By changing the barrier gate voltage we can control the tunnel coupling and find that we can tune the interdot tunnel coupling over a range from completely off to beyond 40 GHz. Note that we can set larger tunnel couplings, see for example Figure \ref{fig:tunnelcoupling}c. However, in this regime we are not able to make reliable fittings of the charge polarization line, due to the reduced charge sensitivity of the sensor, as a result of the merging of $Q_3$ and $Q_4$. 
	
	After focusing on the interdot coupling, we now show that we can form a quadruple quantum dot in the 2x2 array, reaching single-hole occupation for all four quantum dots simultaneously. With both sensors we can detect charge transitions of each quantum dot within the array, although a significantly stronger sensitivity is obtained for the quantum dots neighboring the sensor. In order to conveniently tune and demonstrate the single-hole occupation for all quantum dots, another virtual gate set is defined (see Supplementary Information section II), such that the addition lines of all four dots have a distinctive slope. In Figure \ref{fig:4quantum dotCSD_first_try}a and b we show the charge stability diagram as measured by the individual charge sensors. Taking the derivative of the signal and summing them results in Figure \ref{fig:4quantum dotCSD_first_try}c. The observed charge addition lines are explained in Figure \ref{fig:4quantum dotCSD_first_try}d.        
	
	In conclusion, we have demonstrated shell filling, tunable interdot coupling, and the tuning of a quadruple quantum dot to the single-hole states. The shell filling experiments underpin the high-quality of planar germanium quantum dots as a platform for spin qubits. Moreover, this statement is supported by the demonstration that the tunnel coupling between single holes can be tuned over a large range, from isolated quantum dots to strongly coupled and merging quantum dots. This tunability is promising for quantum simulation with quantum dots such as simulating metal-insulator transitions \cite{Hensgens2017}. Simultaneously, the ability to turn the exchange interaction on and off is highly advantageous for digital quantum computation and can be used to program two-qubit logic at their sweet spots. The demonstration of a quadruple quantum dot positioned in a two-dimensional array  is an important stepping stone toward quantum information processing using standard semiconductor manufacturing.

\section*{Acknowledgements}
We thank Caroline Smulders and all the members of the Veldhorst group for inspiring discussions. M. V. acknowledges support through projectruimte and Vidi grants, associated with the Netherlands Organization of Scientific Research (NWO).

\section*{Competing Interests}
The authors declare no competing interests. Correspondence should be addressed to M.V. (M.Veldhorst@tudelft.nl).    

\clearpage

 \providecommand{\latin}[1]{#1}
\makeatletter
\providecommand{\doi}
  {\begingroup\let\do\@makeother\dospecials
  \catcode`\{=1 \catcode`\}=2 \doi@aux}
\providecommand{\doi@aux}[1]{\endgroup\texttt{#1}}
\makeatother
\providecommand*\mcitethebibliography{\thebibliography}
\csname @ifundefined\endcsname{endmcitethebibliography}
  {\let\endmcitethebibliography\endthebibliography}{}

\end{document}